\documentclass[aps,prb,twocolumn,superscriptaddress,showkeys]{revtex4-2}
\usepackage[hypertexnames=true]{hyperref}
\usepackage{textcomp,gensymb}
\usepackage{graphicx}
\usepackage{lipsum}

\begin{document}
	
\title{Exploring the Anomalous Nernst Effect in SrRuO$_3$}

\author{Anna Merin Francis}
\affiliation{Department of Physics, Indian Institute of Science Education and Research, Pune, India}
\author{Avirup De}
\affiliation{Department of Physics, Indian Institute of Science Education and Research, Pune, India}
\affiliation{Department of Condensed Matter Physics, Weizmann Institute of Science, Israel}
\author{Abhijit Biswas}
\affiliation{Centre for Energy Science, Indian Institute of Science Education and Research, Pune, India}
\affiliation {Department of Materials Science and NanoEngineering, Rice University, Texas, United States}
\author{Lily Mandal}
\affiliation{Centre for Energy Science, Indian Institute of Science Education and Research, Pune, India}
\affiliation{Research Institute for Sustainable Energy (RISE), TCG-CREST, Kolkata, India}
\author{Pallavi Kushwaha}
\affiliation{CSIR-National Physical Laboratory, New Delhi, India}
\author {Sunil Nair}
\affiliation{Department of Physics, Indian Institute of Science Education and Research, Pune, India}
\email{sunil@iiserpune.ac.in}


\date{\today}

\begin{abstract}
We investigate the anomalous Nernst effect in epitaxial SrRuO$_3$ thin films grown on c-cut Al$_2$O$_3$ substrates, and in a polycrystalline SrRuO$_3$ slab.
Through comprehensive measurements of the transverse thermoelectric response as a function of temperature and magnetic field, we observe a pronounced Nernst signal near $T_c$ in the (111)-oriented SrRuO$_3$ thin films. The strong temperature and nontrivial field dependence underscore the pivotal role of SrRuO$_3$’s magnetic anisotropy in tuning the Berry curvature and, consequently, the anomalous Nernst effect.
 
\end{abstract}


\maketitle

The interplay between spin–orbit coupling (SOC) and time-reversal symmetry breaking in metallic ferromagnets generates a finite Berry curvature in momentum space, which acts as an effective magnetic field on charge carriers.\cite{xiao2010berry,RevModPhys.82.1539,xiao2010berry}. This curvature underlies intrinsic transverse effects such as the anomalous Hall and anomalous Nernst effects, whose magnitude and sign depend sensitively on the Berry curvature near the Fermi level\cite{PhysRevLett.93.206602}. The subtle changes in the electronic structure—caused by symmetry breaking, strain, or magnetic anisotropy—can reshape this curvature landscape\cite{PhysRevX.15.031006,vzelezny2023high}, providing a powerful way to tune magnetotransport in correlated ferromagnetic systems. SrRuO$_3$ (SRO), a prototypical correlated 4$d$ itinerant ferromagnet with strong SOC\cite{PhysRevLett.127.256401}, stands as a paradigmatic material for probing Berry-curvature-induced transport anomalies\cite{PhysRevLett.77.2774,PhysRevLett.119.026402,PhysRevMaterials.7.054406}, including potential signatures of magnetic Weyl semimetal states\cite{itoh2016weyl,takiguchi2020quantum}. In bulk form, SRO mainly displays uniaxial magnetocrystalline anisotropy \cite{kanbayasi1976magnetocrystalline}, shaped by the interplay between SOC and its orthorhombic lattice structure. In SRO thin films, however, the anisotropy becomes more tunable\cite{tian2021manipulating, koster2012structure,ning2015anisotropy,wang2022tunable,PhysRevB.85.134429} — shifting between uniaxial (either out-of-plane or in-plane) and biaxial forms, depending on factors such as strain\cite{wakabayashi2021wide,jung2004magnetic}, controlled crystallographic orientation\cite{gan1999lattice,herranz2006controlled}, film thickness\cite{PhysRevResearch.2.032026,jeong2024dimensionality}, and interfacial effects\cite{PhysRevB.88.144412,PhysRevB.94.214420}.\\
Among these, SRO thin films grown along the pseudocubic [111] direction have emerged as a distinct platform, where the trigonal symmetry and tilted octahedral network produce unusual strain responses\cite{tian2021manipulating,lin2021electric}, enabling access to topological phases not present in bulk or conventional (001)-oriented films. In this orientation, strain couples to multiple rotational modes of the RuO$_6$ octahedra, leading to non-collinear Ru–O–Ru bonding and substantial changes in the electronic structure and magnetic anisotropy\cite{PhysRevB.102.041102,ding2023magnetism}. Such symmetry–strain coupling makes the (111)-oriented SRO especially promising for tuning Berry–curvature–driven transport phenomena\cite{cuoco2022materials}. While the extraordinary Hall effect\cite{PhysRevB.70.180407} in SRO has been an area of active investigation over the past decade, its microscopic origin remains debated, invoking a two-channel Hall response \cite{kimbell2020two,PhysRevMaterials.4.104410}, or to a topological Hall contribution arising from noncoplanar spin textures\cite{PhysRevB.102.220406,huang2020detection,roy2023origin}. This contrast highlights the necessity of employing complementary probes to elucidate this behaviour in SRO—especially its thermoelectric analogue, the anomalous Nernst effect (ANE), which has remained largely unexplored.\\ 
\begin{figure}
	\includegraphics[width=1\columnwidth]{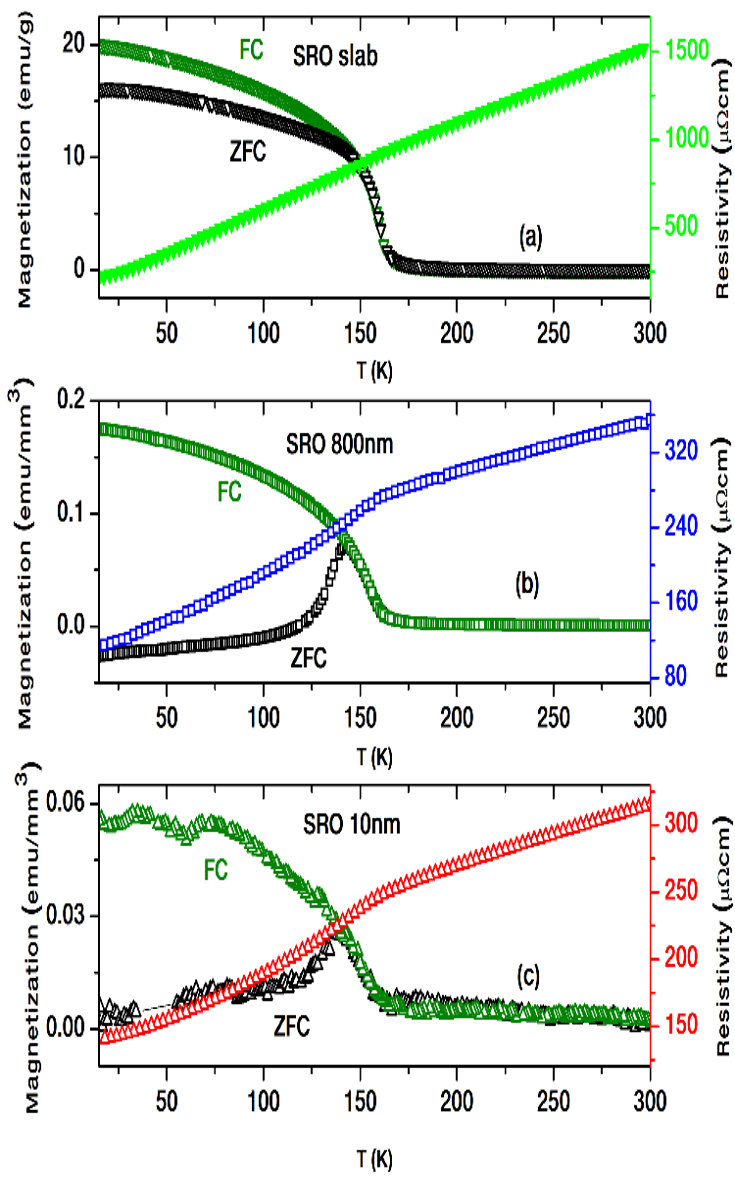}
	\caption{\footnotesize Temperature-dependent resistivity and magnetization measured at an applied inplane magnetic field of 2kOe for (a) polycrystalline SRO slab, (b) SRO 800 nm thin film, and (c) SRO 10 nm thin film, showing distinct Zero-Field-Cooled (ZFC, black) and Field-Cooled (FC, dark green) magnetization curves, indicative of magnetic anisotropy.}
	\label{figure1}
\end{figure}
Our study compares three forms of SRO, specifically a bulk polycrystalline slab (SRO slab), a relatively thick epitaxial film ($\sim$800nm, SRO 800nm), and an ultrathin epitaxial film ($\sim$10nm, SRO 10nm). The polycrystalline SRO slab is fully relaxed and strain-free, exhibiting overall isotropic magnetic behavior. In contrast, the 800 nm epitaxial film retains its crystallinity while relaxing strain, providing a platform to investigate how strain relaxation influences magnetic anisotropy and Berry-curvature–driven responses. Meanwhile, the ultrathin 10 nm film maintains fully coherent strain and exhibits pronounced interfacial effects, enabling the exploration of how confinement and interfacial coupling modulate intrinsic Berry curvature phenomena. The polycrystalline SRO slab was prepared by solid-state synthesis followed by pellet sintering\cite{PhysRevB.58.653}, whereas epitaxial SRO thin films were grown on hexagonal c-cut Al$_2$O$_3$ substrates using pulsed laser deposition (PLD). Despite a larger lattice mismatch (with in-plane strain of $\sim 4.91\%$) compared to other substrates reported for coherent SRO growth, c-cut Al$_2$O$_3$ can still support epitaxial SRO films through optimized deposition techniques, and its potential for magnetotransport and thermoelectric studies remains largely unexplored. Owing to the pseudo-cubic crystal structure of SRO, growth on the hexagonal c-cut Al$_2$O$_3$ substrate naturally favors the (111) orientation, as confirmed by X-ray diffraction (XRD) measurements that reveal (111)-oriented epitaxial growth of the SRO thin films\cite{biswas2020enhanced}. The SRO 800 nm thin film relaxes toward bulk orthorhombic $Pbnm$ symmetry while retaining epitaxial alignment.  The ultrathin  SRO 10 nm layers stay fully coherent with the substrate, shaping the (111) plane into a pseudo-trigonal lattice with threefold rotational symmetry. In case of (001)-oriented SRO thin films grown on substrates such as SrTiO$_3$(001), LSAT(001), and LaAlO$_3$(001), the lattice mismatch is primarily accommodated through biaxial in-plane tensile or compressive strain, resulting in nearly volume-conserving tetragonal or slightly monoclinic distortions\cite{lu2015strain,RevModPhys.84.253}. Similarly, the (110)-oriented thin films deposited on orthorhombic or pseudo-orthorhombic substrates such as NdGaO$_3$(110), GdScO$_3$(110)—experience highly anisotropic in-plane lattice constraints, which can selectively alter octahedral tilt and rotation patterns and stabilize monoclinic or tetragonal structural variants depending on the nature and magnitude of the misfit\cite{kan2013epitaxial}. Unlike (001) and (110)-oriented thin films, (111)-oriented SRO thin films are subject to more complex trigonal strain with strong shear components, making it much more challenging to preserve coherent strain\cite{wang2011magnetic,gan1999lattice,gao2016interfacial}. On SrTiO$_3$(111), ultrathin SRO thin films remain fully coherent and adopt a pseudo-trigonal/rhombohedral-like distortion due to the threefold rotational symmetry imposed by the substrate, whereas on larger-lattice substrates such as KTaO$_3$(111), SRO experiences tensile trigonal strain, often accompanied by octahedral tilt rearrangements\cite{PhysRevB.88.144412,PhysRevB.88.214410}.With increasing thickness, the SRO thin films gradually relax to the bulk orthorhombic $Pbnm$ symmetry.\\
The temperature-dependent resistivity and magnetization measurements confirm the ferromagnetic metallic nature\cite{PhysRevB.63.054435, rastogi2019metal} of all three samples, as shown in FIG.~\ref{figure1}. A continuous paramagnetic-to-ferromagnetic phase transition, with the Curie temperature ($T_c$) reaching $\sim 160$K in the polycrystalline SRO slab, consistent with the value reported for bulk SRO is observed.  In the SRO 800nm film, $T_c$ decreases slightly to $\sim 155$K, consistent with the finite-size effect\cite{PhysRevLett.28.1516}, where $T_c$ scales with the thickness of the ferromagnetic material. By contrast, the SRO 10nm thin film shows a suppression of $T_c$ to $\sim 142$K, arising from the combined influence of substrate-induced strain, interface-driven octahedral distortions, and finite-size effects that hinder the stabilization of long-range ferromagnetic order at reduced dimensionality\cite{lu2015strain,kar2021high}. Notably, the SRO thin films display a pronounced bifurcation between the zero-field-cooled (ZFC) and field-cooled (FC) magnetization curves, whereas the polycrystalline SRO slab exhibits a well-aligned magnetization response along the applied magnetic field direction. Although this behavior has previously been reported in SRO and attributed to magnetic anisotropy \cite{kumar1999irreversible, joy1998relationship}, it is typically observed only at much lower magnetic fields (below 1kOe). In contrast, the clear separation between the ZFC and FC curves observed at 2kOe in our SRO thin films indicates strong magnetocrystalline anisotropy (MCA) \cite{kanbayasi1976magnetocrystalline, PhysRevB.81.184418}.\\
The strength and orientation of the MCA in SRO thin films is strongly influenced by the epitaxial strain imposed by the substrate, the intrinsic crystallographic symmetry, and spin–orbit coupling of the Ru 4$d$ electrons. In orthorhombic SRO, which is typical for (001) or (110)-oriented films, MCA is generally uniaxial, with the magnetic moments preferentially aligned along a single crystallographic direction. In bulk SRO, the magnetic easy axis typically lies along the pseudocubic [001] or [110] directions, depending on the nature and magnitude of structural distortions\cite{PhysRevB.94.214420,wakabayashi2021wide}. While the compressive strain in SRO thin films tends to stabilize an easy axis parallel to the film normal, the tensile strain in SRO thin films favors in-plane magnetization. Under compressive strain, the Ru 4$d$–O 2$p$ orbital overlap along the out-of-plane direction is reduced, which disrupts the quenching of the out-of-plane orbital magnetic moment and leads to a pronounced out-of-plane magnetic easy axis. In contrast, tensile strain is expected to similarly decrease the in-plane orbital overlap, producing an orbital-moment anisotropy that favours an in-plane magnetization direction \cite{PhysRevB.88.214410}. In contrast to the uniaxial anisotropy characteristic of the orthorhombic SRO, the trigonal SRO shows a planar magnetic anisotropy below $T_c$\cite{ding2023magnetism,PhysRevB.81.184418}, with its magnetic moments oriented within the (111) plane.\\
The strong SOC in SRO introduces relativistic corrections that render the magnetic interactions direction-dependent. 
Within the Hamiltonian formalism, SOC is expressed as $ H_{\rm SOC} = \lambda \sum_i \mathbf{L}_i \cdot \mathbf{S}_i$, where $\mathbf{L}_i$ and $\mathbf{S}_i$ denote the orbital and spin angular momenta of the $i$-th Ru ion, and $\lambda$ is the SOC constant. Through this interaction, SOC lifts the degeneracy among possible magnetic configurations and dictates the preferred spin orientation \cite{PhysRevB.107.174429}.
\begin{figure}
	\includegraphics[width=1\columnwidth]{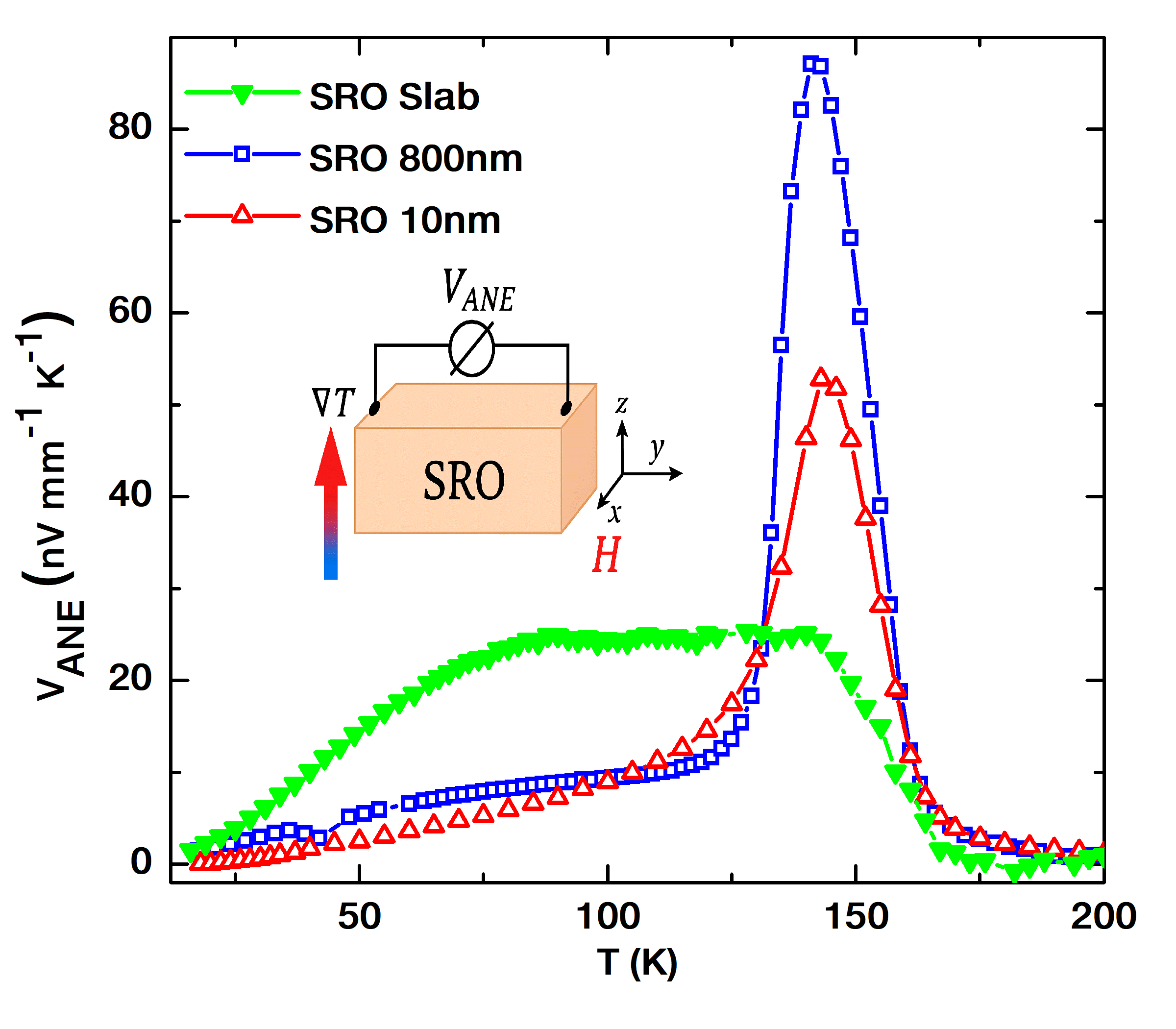}
	\caption{\footnotesize Temperature-dependent anomalous Nernst effect (ANE) measurements on (a) polycrystalline SRO slab, (b) SRO 800nm thin film and (c) SRO 10nm thin film. The ANE was measured under an applied magnetic field of 2kOe and a thermal gradient of 10K and the corresponding schematic of the measurement is illustrated in the inset. A pronounced Nernst signal near $T_c$ is observed in the (111)-oriented SRO thin films.}
	\label{figure2}
\end{figure}
During ZFC, the domains remain randomly oriented and frozen at low T. Upon warming, thermal energy overcomes the pinning and anisotropy barriers, allowing spins to reorient along the applied field direction and leading to an increase in magnetization near $T_c$. In addition to the pronounced divergence between ZFC and FC magnetization, a negative magnetization is observed in the ZFC measurements at low temperatures for the SRO 800nm thin film. This behavior does not arise from the residual trapped fields in the magnetometer, but is consistent with the negative magnetization observed in ZFC measurements of SRO reported in previous studies\cite{sarkar2013temperature, PhysRevLett.85.5182}. During ZFC, certain domains align along local easy axes that are antiparallel to the applied field, contributing a negative component to the net magnetization. When such domains dominate at low temperatures\cite{marshall1999lorentz}, the resulting ZFC magnetization can appear negative.\\
To study the Berry-curvature response, comprehensive temperature and magnetic field-dependent ANE measurements were performed on all three SRO samples under an applied magnetic field of 2kOe and a thermal gradient of 10K\cite{ghosh2018large,PhysRevB.103.L020404}, as depicted in the inset of FIG.~\ref{figure2}. The anomalous Nernst voltage develops perpendicular to both the magnetization and the heat-flow direction and is given by,
$V_{T} = S_{xy} \left( \mathbf{M} \times {\nabla}T \right)$, where $S_{xy}$ is the transverse Seebeck coefficient. In the polycrystalline SRO slab, the temperature dependence of the ANE closely follows the temperature dependence of the transverse Peltier coefficient $\alpha_{xy}$ reported for bulk SRO single crystals \cite{PhysRevLett.99.086602}. The 
$\alpha_{xy}$ rises sharply just below $T_c$, remains nearly proportional to $M$, decreases approximately linearly at low temperatures, and eventually vanishes as $T \rightarrow 0$. While ANE is the converse of the anomalous Peltier effect, related via Onsager reciprocity\cite{PhysRevLett.131.246302}, the Peltier coefficient can be expressed using the Mott's relation as $\alpha_{xy} = \frac{\pi^2}{3} \frac{k_B^2 T}{e} \frac{d\sigma_{xy}}{d\epsilon}$, where $\sigma_{xy}$ is the transverse (Hall) conductivity, and the derivative $\frac{d\sigma_{xy}}{d\epsilon}$ reflects the evolution of the band structure at the Fermi level\cite{PhysRevLett.97.026603}. The behavior of bulk SRO can be understood from the above formula, where just below $T_c$, $\alpha_{xy}$ is determined by 
$\frac{d\sigma_{xy}}{d\epsilon}$, reflecting changes in the Fermi-level band structure arising from the ferromagnetic transition.\\
Similarly, in the case of SRO thin films, the transverse Nernst signal shows a monotonic increase with temperature up to $T_c$, without any reversal of polarity unlike the AHE; instead, it follows a clear scaling trend reminiscent of the ZFC magnetization. The AHE is determined by the total Berry curvature integrated over all occupied states and thus reflects the global topology of bands below the Fermi energy $E_f$ \cite{RevModPhys.82.1539}, while the ANE is governed by the Berry curvature of states at or near the Fermi level, weighted by $E_f$\cite{PhysRevB.96.174406}. The ANE thus directly probes the electronic structure and Berry curvature at the Fermi surface, rather than the full occupied band manifold. This makes it particularly sensitive to the local topological features near $E_f$\cite{narang2021topology}. In SRO, strong Ru 4$d$–O 2$p$ hybridization gives rise to metallic bands, with $E_F$ passing through the degenerate $t_{2g}$ orbitals. The SOC is predicted to lift degeneracies within these $t_{2g}$ states, creating an anisotropic magnetic energy landscape, giving rise to a non-zero Berry curvature $\Omega(\mathbf{k})$ near $E_F$ \cite{tian2021manipulating,huang2021first}. The enhanced ANE signal in the SRO 800nm thin film compared to the SRO 10nm thin film is also shaped by the density of states, carrier scattering, and both the magnitude and orientation of magnetization. In the polycrystalline SRO slab, the reduced ANE arises from lower magnetization, averaging of magnetic anisotropy due to randomly oriented grains, and additional attenuation from grain boundary scattering and structural defects.\\
\begin{figure}
	\includegraphics[width=1\columnwidth]{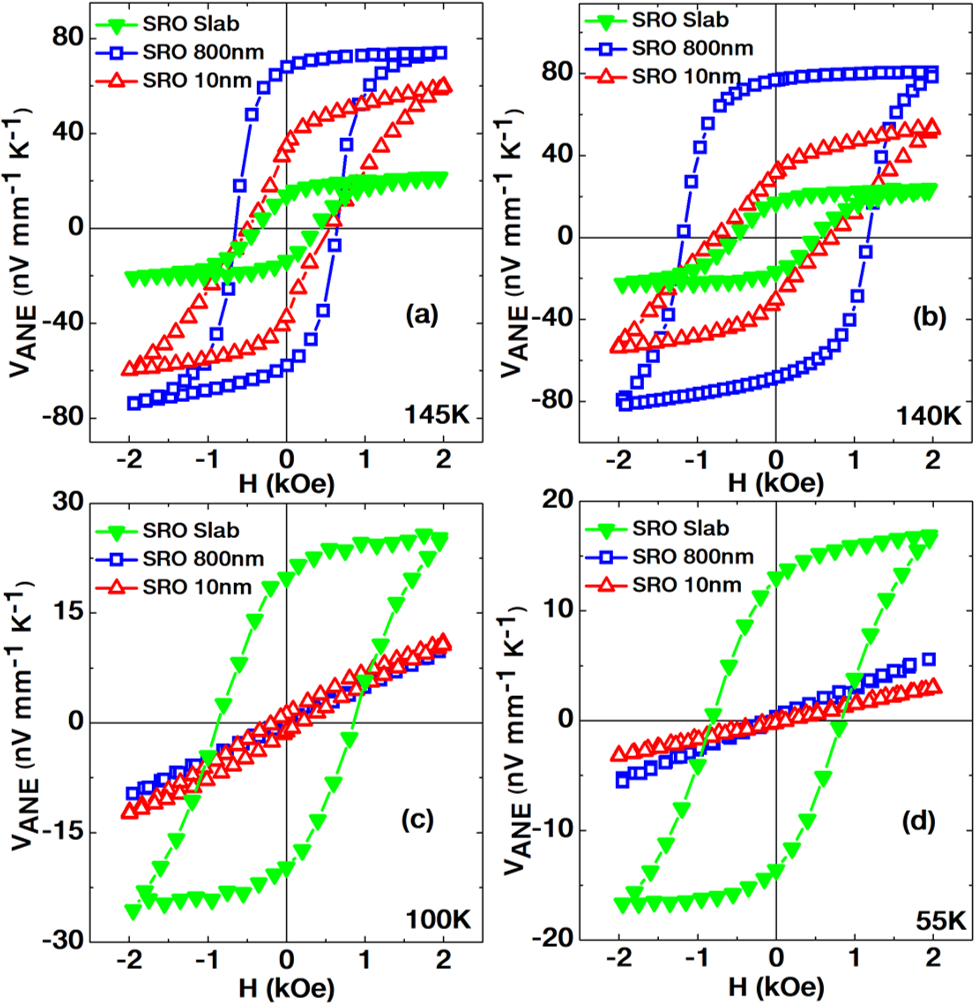}
	\caption{\footnotesize Magnetic field-dependent-anomalous Nernst effect measurements on polycrystalline SRO slab along with SRO 800nm and SRO 10nm thin films grown on c-cut Al$_2$O$_3$ substrates, under a temperature gradient of 10K at various temperatures. A clear variation in loop area is observed, where the SRO 800nm and SRO 10nm thin films display pronounced broadening near 
    $T_c$, whereas the polycrystalline sample shows a larger loop area at 55 K.}
	\label{figure3} 
\end{figure}
\begin{figure}
	\includegraphics[width=1\columnwidth]{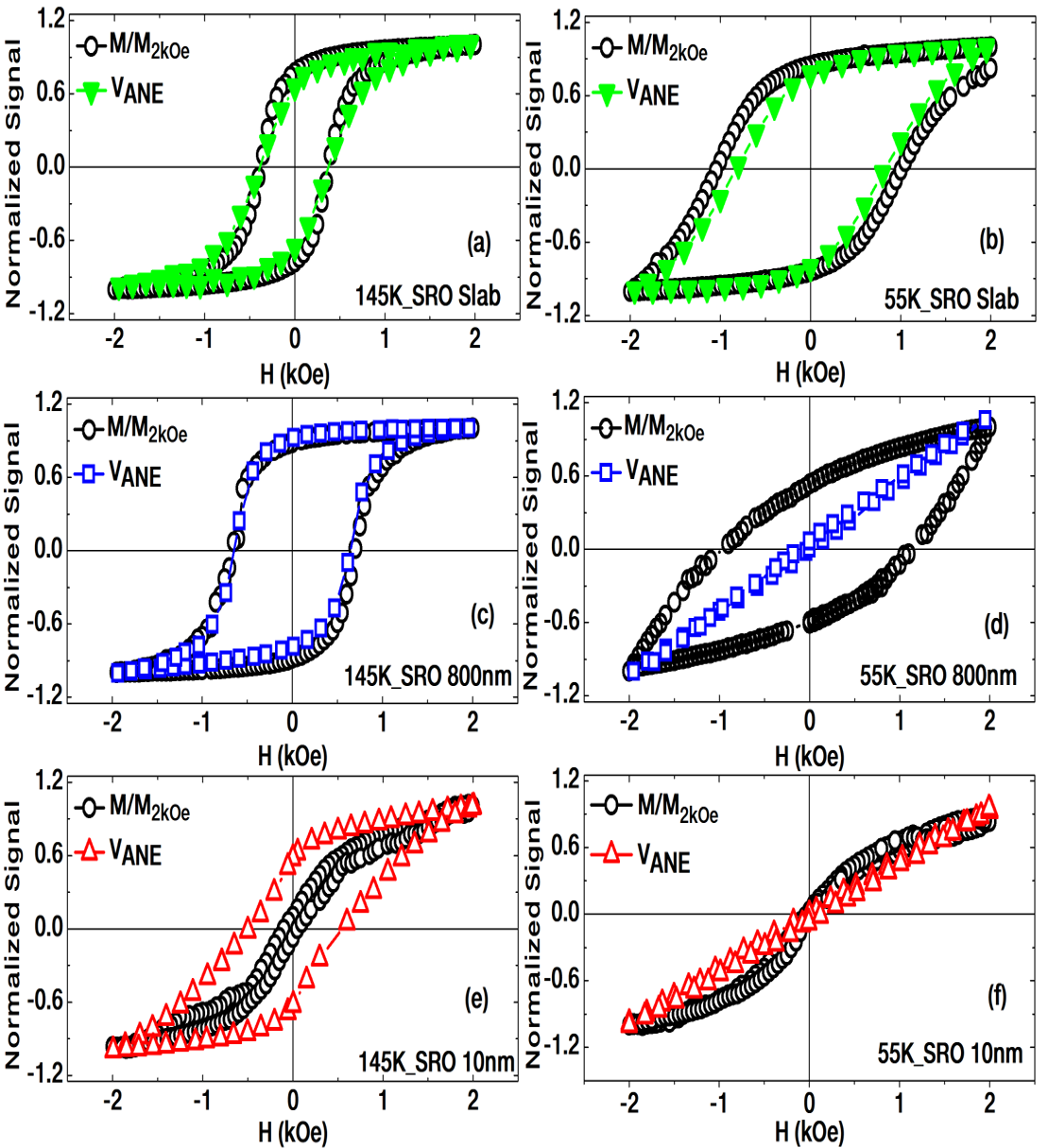}
	\caption{\footnotesize Comparison of magnetic hysteresis loops and Field-dependent anomalous Nernst effect measured at 145K and 55K on SRO slab alongside SRO 800nm and SRO 10nm thin films on c-cut Al$_2$O$_3$ substrates, under an applied thermal gradient of 10K. At 55K, both SRO 800nm and SRO 10nm thin films exhibited a similar trend regardless of their magnetic hysteresis.}
	\label{figure4}
\end{figure}
To further investigate the Nernst response in SRO, we performed magnetic-field-dependent ANE measurements on all three samples across a range of temperatures. As illustrated in FIG.~\ref{figure3}, near $T_c$, the SRO slab exhibits a lower ANE signal than both SRO thin films, whereas the SRO 800nm thin film shows a higher signal than the SRO  10nm thin film. At 55K, the SRO slab displays a pronounced ANE signal compared to the SRO thin films, reflecting their temperature-dependent Nernst behavior. In addition to the Nernst signal strength, a clear variation in the Nernst loop area is observed; near $T_c$, the SRO 800 nm film shows a larger loop area, whereas at 55 K, the SRO slab exhibits a larger area, and both the SRO 800 nm and SRO 10 nm thin films display similar area. The Nernst loop area is sensitive to the strength of the Berry curvature near the Fermi level, which in turn depends on the material’s electronic structure and magnetization\cite{PhysRevB.84.174439}. To examine how the ANE varies with respect to magnetization, we systematically compared the field-dependent Nernst response at 145K and 55K alongside the magnetic hysteresis, as shown in FIG.~\ref{figure4}. In ANE measurements, conventional skew scattering scales approximately linearly with the magnetization, as the asymmetric scattering of charge carriers is directly linked to the net magnetic moment. The anomalous Hall conductivity associated with skew scattering can be expressed as $\sigma_{xy}^{\rm skew} \propto M \cdot \rho_{xx}^{-1}$, where $\rho_{xx}$ is the longitudinal resistivity, governed by scattering rates\cite{PhysRevB.84.174439,xiao2010berry}. In the SRO slab, the Nernst response is observed to scale directly with magnetization at both 145K and 55K, indicating that it is predominantly governed by extrinsic skew scattering. At 145 K, the SRO 800 nm thin film shows a clear scaling between the Nernst loop and the magnetization, indicating that skew scattering dominates, whereas at 55 K it develops a noticeable change in the Nernst loop area relative to its magnetic hysteresis. In contrast, the SRO 10nm thin film deviates from the magnetization behavior at both temperatures. Unlike skew scattering, which scales linearly with $M$, intrinsic contributions can exhibit a nonlinear or more complex dependence on $M$\cite{guin2019anomalous}. The variation of the Nernst loop area with respect to magnetization observed in both the SRO 800nm and SRO 10nm thin films suggests that this type of ANE behavior occurs only when the intrinsic momentum-space Berry curvature contributes.\\ 
In SRO thin films with strong MCA, the Berry curvature distribution is more rigidly aligned along the easy axis, so variations in $M$—caused by changes in temperature or magnetic field—affect the intrinsic contribution in a predictable yet nonlinear manner. The anomalous behavior observed at 55K in the SRO 800nm thin film—both relative to its magnetization and in comparison with the SRO 10nm thin film at the same temperature—suggests an intrinsic Berry-curvature contribution, possibly of topological origin. Topological features such as Weyl nodes—emerging near the Fermi level due to exchange splitting and strong spin–orbit coupling, which enhance the Berry curvature contributions \cite{liu2018giant,PhysRevB.108.075164}—can play a crucial role in the intrinsic anomalous Nernst effect (ANE). Although Weyl nodes have been reported to contribute to the intrinsic anomalous Hall effect in (111)-oriented SRO \cite{ding2023magnetism,lin2021electric}, the topological origin of the effects observed in the SRO thin films studied here requires further investigation. The differences between the Nernst loop and magnetic hysteresis in the SRO 800nm and SRO 10nm thin films can be attributed to their distinct magnetic anisotropies—uniaxial in the SRO 800nm film and planar in the 10nm film. Even in the absence of Weyl points, magnetic anisotropy can still tune the Berry curvature by governing the magnetization alignment, SOC, and overall band structure\cite{tian2021manipulating}.
\\
Although the origin of the extraordinary Hall effect in SRO thin films remains unresolved, the ANE measurements presented here demonstrate that magnetic anisotropy plays a decisive role. The strong magnetic anisotropy restricts the permissible canting or tilting of spins, thereby suppressing the formation of topologically non-trivial spin configurations unless the energetics along specific crystallographic directions are favorable. Consequently, chiral spin textures—such as skyrmions—can stabilize only when the Dzyaloshinskii–Moriya interaction (DMI) energy becomes comparable to or surpasses the magnetic anisotropy energy barrier. Under such conditions, and within the limits imposed by the applied magnetic field\cite{PhysRevResearch.2.032026}, a topological Hall contribution may be observable. To the extent reported in the literature, this comprehensive ANE study on polycrystalline SRO alongside epitaxial thin films provides critical insights into Berry-curvature engineering in SRO, highlighting the pivotal role of magnetic anisotropy in modulating transverse thermoelectric transport.\\ 
In conclusion, the temperature- and magnetic-field-dependent ANE measurements demonstrate a marked amplification of the Nernst response near ($T_c$) in the (111)-oriented SRO thin films. The comparative analysis across the three distinct SRO samples highlights that the ANE is highly responsive to variations in magnetic anisotropy, which in turn modulates the Berry curvature. The unusual behavior exhibited by the (111)-oriented films further emphasizes how crystallographic orientation—and the accompanying strain state—critically influences Berry-curvature–driven transport. These findings establish the ANE as an effective probe for exploring Berry curvature and topological effects in itinerant ferromagnets.\\
A.M.F. acknowledges Prof. Ogale for his guidance in thin film growth. A.M.F. thanks Murthykrishnan and Harshdip for their assistance with the ANE measurements on the SRO slab, and R. Godbole for his support during the XRR measurements. A.M.F. acknowledges DST-INSPIRE for providing financial support through Senior Research Fellowship (SRF). A.M.F. and S.N. acknowledge the funding support by the  Department of Science and Technology (DST, Government of India), through grant number CRG/2022/003316.
\bibliography{manuscript}

\end{document}